\documentclass[twocolumn]{aastex631}
\usepackage{amsmath}
\shorttitle{Devouring the Milky Way Satellites}
\shortauthors{Weerasooriya et al.}
\graphicspath{{./}{figures/}}
\usepackage{comment}
\begin{document}


\title{Devouring the Milky Way Satellites: Modeling Dwarf Galaxies with Galacticus}

\author[0000-0001-9485-6536]{Sachi Weerasooriya}
\author[0000-0003-4037-5360]{Mia Sauda Bovill}
\affiliation{Department of Physics and Astronomy, Texas Christian University, Fort Worth, TX 76109, USA}

\author[0000-0001-5501-6008]{Andrew Benson}
\affiliation{Carnegie Observatories, 813 Santa Barbara Street, Pasadena, California, 91101 USA}

\author{Alexi M. Musick}
\affiliation{
Department of Physics and Astronomy, University of Oklahoma, Norman, OK 73019}

\author[0000-0003-4223-7324]{Massimo Ricotti}
\affiliation{Department of Astronomy, University of Maryland, College Park, MD 20742}

\begin{abstract}

Dwarf galaxies are ubiquitous throughout the universe and are extremely sensitive to various forms of internal and external feedback. Over the last two decades the census of dwarf galaxies in the Local Group and beyond has increased markedly. While hydrodynamic simulations (e.g. FIRE II, MINT Justice League) have reproduced the observed dwarf properties down to the ultra-faints, such simulations require extensive computational resources to run. In this work, we constrain the standard physical implementations in the semi-analytic model Galacticus to reproduce the observed properties of the Milky Way satellites down to the ultra-faint dwarfs found in SDSS.  We run Galacticus on merger trees from our a high resolution N-body simulation of a Milky Way analog. We determine the best fit parameters by matching the cumulative luminosity function and luminosity-metallicity relation from both observations and hydrodynamic simulations. With the correct parameters, the standard physics in Galacticus can reproduce the observed luminosity function and luminosity-metallicity relation the Milky Way dwarfs. In addition, we find a multi-dimensioinal match with half-light radii, velocity dispersions and mass to light ratios at $z~=~0$ down to $M_V \leq -6$ ($L \geq 10^4$~L$_\odot$). In addition to successfully reproducing the properties of the $z~=~0$ Milky Way satellite population, our modeled dwarfs have star formation histories which are consistent with those of the Local Group dwarfs.




\end{abstract}
\keywords{Dwarf galaxies (416) --- Galaxy evolution (594) --- Galaxy formation (595) --- Galaxy quenching (2040) --- Theoretical models (2107)}
\section{Introduction}\label{sec:intro}

Dwarf galaxies are the most common type of galaxy in the universe \citep{ferguson1994} and their shallow potential wells make them extremely sensitive to internal and external feedback \citep[{\it e.g.},][]{dekel1986,thoul1996,benson2002a,okamoto2010}. Thus, they are excellent probes of both dark matter \citep[{\it e.g.},][]{PolisenskyR:2011} and internal/external environmental physical processes. 

In the last fifteen years, the census of dwarf galaxies in the Local Group has more than doubled due to the SDSS \citep{sdss2014,SDSS}, the DES \citep{DES}, and PAndAS \citep{pandas2016}. In addition, ongoing and upcoming surveys are pushing the frontiers of observational studies of dwarf galaxies beyond the Local Group. These include, among others, targeted surveys of Centaurus A \citep{SCABS, PISCES}, M81/M82 \citep{sorgho2019}, and wider surveys such as SAGA \citep{geha2017,mao2021}. 

In concert with our increased understanding of the observational properties of dwarf galaxies, theoretical studies of the fossils of the first galaxies \citep{BovillR:2009, RicottiPG:2016,Wheeleretal:2019} and of dwarf satellites around the Milky Way using hydrodynamical simulations have come of age \citep{Wetzel+2022,Applebaum2021}. FIRE-II \citep{FIRE2018, Wetzel+2022} simulates a Milky Way analog resolving star formation physics down to low mass dwarfs ($M_V<-8$). The MINT Justice League simulations have reproduced properties of the Milky Way satellite system, including the half-light radii, velocity dispersion, and metallicity of the ultra-faints \citep{Applebaum2021}.

Despite their success, high resolution hydrodynamical simulations of Milky Way analogs require extensive computational resources. As our observational sample of dwarf galaxies expands beyond the Local Group, there is a need to simulate dwarf satellite systems in a wider range of environments than can be currently explored by hydrodynamical simulations. Moreover, exploration of the astrophysics involved in any part of the baryon cycle is crucial to understanding how sensitive the feedback mechanisms in dwarf galaxies are to their local environment. Due to their high computational costs, hydrodynamical simulations are not the ideal tool for this work.

In semi-analytic models (SAMs) the baryonic physics is approximated by a set of interconnected differential equations to model the baryonic evolution of galaxies through cosmic time. This is an efficient way of modeling galaxies \citep[e.g.][]{Henriques2009, BensonBower2010,Bower2010} and permits rapid modeling of dwarf galaxies in a range of environments.

Various studies have used SAMs to model galaxies. These analytic models were first proposed by \cite{white1978}, and advanced by \cite{whiteFrenk1991}, \cite{Kauffmann1993}, \cite{somerville1999}, \cite{Cole2000}, \cite{Hatton2003}, \cite{Monaco2007}, \cite{somerville2015}, and others. While the first SAMs were built on Extended Press-Schechter (EPS) merger trees \citep{press1974}, they can now be applied to merger trees from N-body simulations \citep[e.g.][]{Kauffmann1999,Helly2003}.

\cite{sommerville2020} have tested the Santa-Cruz SAM \citep{somerville1999} against the FIRE-II cosmological simulations. Although their stellar-halo mass relations and stellar mass assembly histories agree well with FIRE-II, ISM masses agree only for higher mass halos. In order to reproduce gas accretion efficiencies of FIRE-II dwarfs, they implement a mass dependent preventative feedback model to suppress accretion of gas into halos. Note that ``preventative feedback'' here means preventing accretion of gas onto halos via stellar feedback \citep{Lu2017,sommerville2020}. However, details between implementations in the Santa Cruz SAM \citep{sommerville2020} and the SAM by \cite{Lu2017} vary.

In this work, we will determine whether the Galacticus \citep{galacticus}, run on high resolution merger trees from a cosmological N-body simulation, can reproduce the properties of the Milky Way dwarfs.

In Section \ref{sec:sim} we describe our simulation and the process of constraining Galacticus. In Section \ref{sec:results}, we present predictions from the constrained Galacticus parameters down to ultra faint dwarf scales. Next, we discuss our results and limitations in Section \ref{sec:discussion}, and finally summarize our findings in Section \ref{sec:summary}.

\section{Simulations}\label{sec:sim}

We run an  N-body simulation of a Milky Way analog from $z~=~150$ to $z~=~0$ with WMAP9 cosmology ($\sigma_8\sim0.821,\,H_0\sim\,70.0 km/s/Mpc,\,\Omega_b\sim0.0463,\,\Omega_{\Lambda}\sim0.721$). Initial conditions were generated with MUSIC \citep{Hahn2011} and the simulation was run with Gadget~2 \citep{gadget}. The simulation is analyzed with both the AMIGA \citep{AHF} and Rockstar \citep{rockstar} halo finders. Merger trees are generated using the Consistent Trees \citep{consistent_trees}.

We select an isolated Milky Way analog from a $50\;\hbox{Mpc}\,h^{-1}$ box with $N_\mathrm{eff}~=~256^3$ run from $z~=~150$ to $z~=~0$, resolving the Milky Way candidates at $z\,=\,0$ with $N~>~1000$ particles. Our isolation criteria is a $\sim10^{12} M_\odot$ halo, with no halos greater than $10^{12}\,M_{\odot}$ within $3\,\hbox{Mpc}\,h^{-1}$ at $z\,=\,0$. We select a MW analog with $1.8\times10^{12}\;M_{\odot}$ ($1.2\times10^{12}\;M_{\odot}/h$) following the above conditions. The high resolution region at $z~=~150$ is defined by the particles within $5~R_\mathrm{vir}$ from  the Milky Way analog at $z~=~0$. The highest resolution region has $N_\mathrm{eff}~=~4096^3$, resolving $10^7~M_\odot$ halos with at least 100 particles and softening of $\epsilon=200\,pc$ (physical units).

All simulations described in this work  were run on the Maryland High Performance Computer Cluster Deepthought 2 \footnote{\protect\href{http://hpcc.umd.edu}{\tt http://hpcc.umd.edu}}).

 =We then run Galacticus on the resulting merger trees. Note that most dark matter halo properties (total mass, NFW scale length) used in Galacticus are preset from the N-body trees with the exception of halo spins. Halo spins are typically not well-measured in halos with fewer than of order 40,000 particles \citep{2017MNRAS.471.2871B}.



    
    




\subsection{Constraining Galacticus with the Milky Way Satellites}\label{sec:sam}

We determine the set of Galacticus' parameters which best fit the observed luminosity function and the luminosity metallicity relation for dwarf satellites of the Milky Way. We compare the galaxy models to the updated \cite{mcconnachie2012} table as of Jan 2021 \footnote{\protect\href{https://www.cadc-ccda.hia-iha.nrc-cnrc.gc.ca/en/\\community/nearby/}{\tt https://www.cadc-ccda.hia-iha.nrc-cnrc.gc.ca/en/\\community/nearby/}}. In addition, we have added few satellites from \cite{drlica-wagner2020} that are missing from \cite{mcconnachie2012}. Note, we do not do any formal fitting. Instead, we run a grid of models and choose the ones which produce the best match based on a ``by-eye" judgement. We start from Galacticus' standard set of parameters\footnote{\protect\href{https://github.com/galacticusorg/galacticus/blob/889ab5d347001c9623d74609b51850c080829f96/parameters/baryonicPhysicsConstrained.xml}{\tt github.com/galacticusorg/galacticus/blob/\\889ab5d347001c9623d74609b51850c080829f96/parameters/\\baryonicPhysicsConstrained.xml}} constrained to match the baryonic physics of massive galaxies. Unless mentioned below, we use the parameters given in the file above.

The parameters for massive galaxies have been calibrated to observational datasets, including the stellar mass halo relation of \cite{leauthaud2012} and its scatter from \cite{more2009}, the $z~<~0.06$ stellar mass function of galaxies from the GAMA survey \citep{baldry2012},  the $z~=~2.5-3.0$ stellar mass functions of galaxies from the ULTRAVISTA survey \citep{muzzin2013}, the $z\,=\, 0$ HI mass function of galaxies from the ALFALFA survey \citep{martin2010}, the $z\,=\,0$ black hole mass-bulge mass relation of \cite{kormendy2013}, size distributions of SDSS galaxies from \cite{shen2003},  $H\alpha$ luminosity functions from HiZELS \citep{sobral2013}, GAMA \citep{gunawardhana2013}, g and r-band luminosity functions of SDSS galaxies \citep{montero-dorta2009}, the gas-phase mass-metallicity relation \citep{blanc2019}, and the morphological fraction as a function of stellar mass from GAMA \citep{moffett2016}.

\begin{figure}[h!]
\centering
\includegraphics[width=8cm]{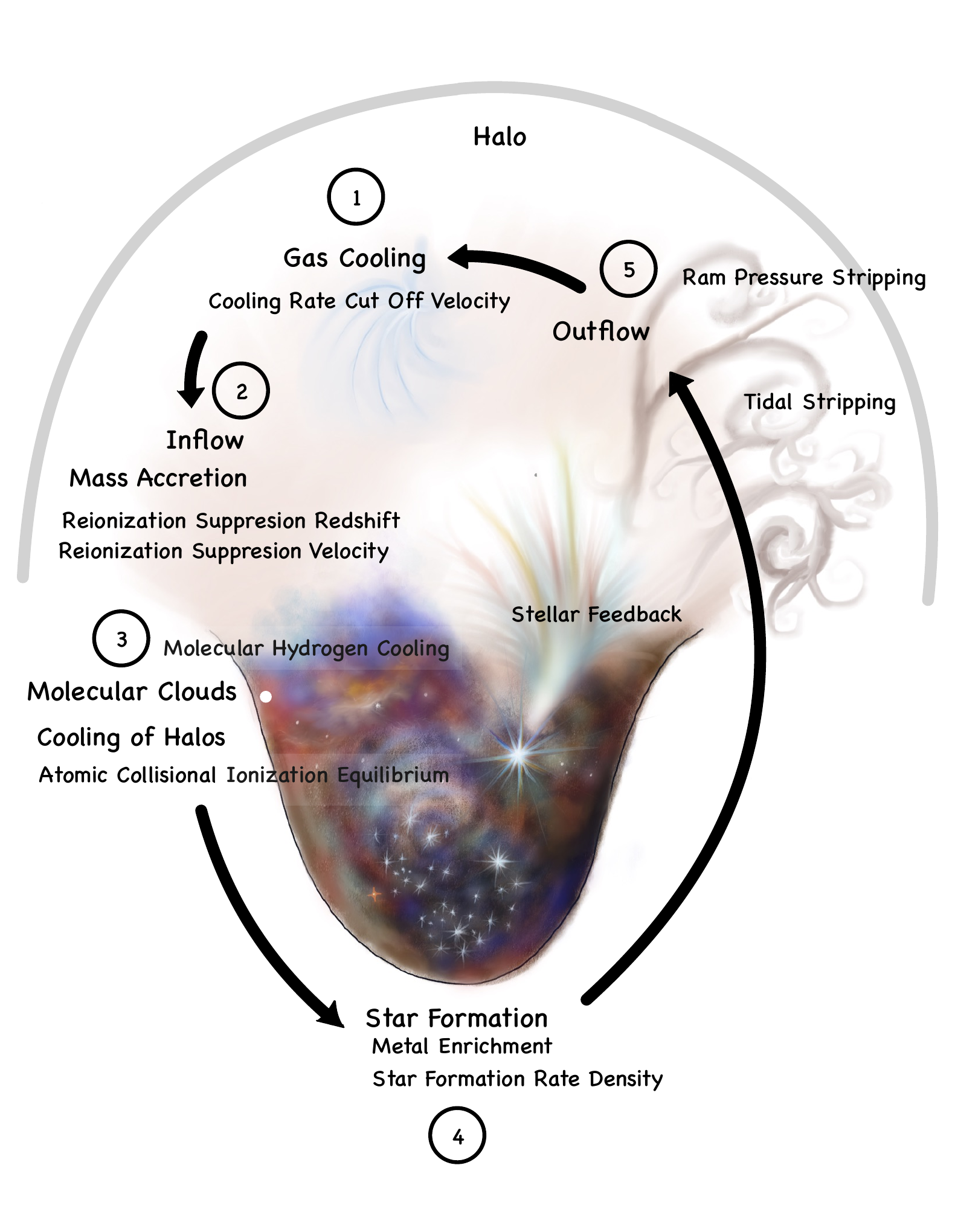}
\caption{Diagram of the baryon cycle. Stage~1: cooling of gas inside the hot halo. Stage~2: cooled gas flows into the halo. Stage~3: the accreted material then forms gas clouds. These gas clouds then start cooling due to molecular hydrogen and collisional ionization equilibrium. Stage~4: cooling of gas aids in the process of star formation. Stage~5: finally, some of the material inside the halo flows back into the hot halo due to feedback effects such as tidal stripping, ram pressure stripping, and supernovae.}
\label{fig:baryon_cycle}
\end{figure}

In the following sections we discuss how we systematically modify the parameters to optimize the baryon cycle (Figure \ref{fig:baryon_cycle}) and reproduce the observed luminosities and metallicities of the Milky Way satellites. We divide our discussion of the modified parameters into two subsets, those that are well-constrained by astrophysics governing dwarf galaxies or their properties (Section~\ref{subsec:constrained}), and those that are not (Section~\ref{subsec:unconstrained}). 

\subsubsection{Parameters with Astrophysical Priors}\label{subsec:constrained}

We begin with parameters whose values are determined, or at least limited, by the astrophysics governing dwarf galaxies or the derived properties of the Milky Way.\\

\noindent{\it {Cooling velocity:}} Atomic hydrogen cooling is suppressed for low mass halos with virial temperatures below $\sim\,10^4\;\hbox{K}$. This is known as the atomic hydrogen cooling limit which corresponds to a virial velocity of $\sim16$ km/s \citep{fitts2017,graus2019}. To suppress star formation in the least massive halos, Galacticus uses a minimum $v_\mathrm{vir}$ below which gas in a halo will be unable to cool and form stars ($v_\mathrm{cooling}$). In this work, we choose $v_\mathrm{cooling}$ values to approximate this atomic cooling limit since gas accretion onto and star formation in halos below the atomic cooling limit is inefficient. Similar thresholds have been used in several high resolution hydrodynamic simulations \citep{sawala2016,munshi2017,benitez-Llambay2017,fitts2017,maccio2017}. Note that the Collisional Ionization Equilibrium (CIE) cooling function does not drop entirely to zero below this threshold due to contributions from metal cooling.  Modeling star formation in halos below the atomic cooling threshold requires accounting for the stochastic effects of $H_2$ cooling, and is beyond the scope of the current work. Therefore, we only consider halos which are above the atomic cooling limit and narrow our choices of velocities ($v_\mathrm{cooling} = 15~-~20$~km~$s^{-1}$).\\

\noindent{\it {Reionization redshift:}} The redshift of reionization for the Milky Way and its local environment is set at $z_\mathrm{reion}~=~9$. This value falls within the range of reionization redshifts calculated by previous works \citep{gnedin2000,bullock2000,alvarez2009,busha2010,iliev2011,spitler2012,ocvirk2013,li2014L,aubert2018}.\\

\noindent{\it{Filtering velocity:}} During and after reionization the reheating of the IGM suppresses the accretion of gas onto low mass halos below the filtering mass \citep{RicottiGnedin:05}. Galacticus parameterizes this with the reionization suppression velocity $v_\mathrm{filter}$. Therefore, accretion of gas is suppressed in halos with $v_\mathrm{virial} \leq v_\mathrm{filter}$. When modelling the effects of reionization on halos across a range of redshifts, this criterion is the superior choice (compared to, for example, a halo mass-based criterion) as the virial velocity of a halo is a direct and redshift-independent measure of the depth of the potential well.\\

\noindent\textit{{Star formation law in disks:}} We calculate the star formation rate density for the disks using the model of \cite{blitz2006}. We choose this prescription because it is based on the astrophysics of molecular hydrogen as opposed to fits of observed data from more massive galaxies \citep{kennicutt1998,shi2011}. This method describes a star formation prescription based on hydrostatic pressure. It uses the linear relation between pressure and the ratio of molecular to atomic gas in galaxies. The stellar surface density rate is computed by
\begin{equation}
    {\dot \Sigma_*}(R)=\nu_\mathrm{SF}(R)\Sigma_\mathrm{H_2,disk}(R).
    \label{eq:sfDisk}
\end{equation}
Here the star formation frequency is given by $\nu_\mathrm{SF}(R)=\nu_\mathrm{SF,0}\left[1+{\left(\frac{\Sigma_\mathrm{HI}}{\Sigma_0}\right)}^q\right]$, where  $\Sigma_0$ is the critical surface density for formation of molecules and $q$ is an exponent. Note that the star formation efficiency is suppressed in `subcritical' regions where $\Sigma_\mathrm{HI}<\Sigma_0$.
The surface density of molecular gas is given by
\begin{equation}
    \Sigma_\mathrm{H_2}={\Big(\frac{P_\mathrm{ext}}{P_0}\Big)}^{\alpha}\Sigma_\mathrm{HI},
\end{equation}
 where $P_0$ is the characteristic pressure and $\alpha$ is the pressure exponent (we use $\alpha=0.92$ as suggested by \citet{blitz2006}). External hydro-static pressure within a gas cloud in the disk is calculated by,
 \begin{equation}
     P_\mathrm{ext}=\frac{\pi}{4}G\Sigma_\mathrm{gas}\Bigg[\Sigma_\mathrm{gas}+\Big(\frac{\sigma_\mathrm{gas}}{\sigma_*}\Big)\Sigma_*\Bigg],
 \end{equation}

\noindent where $\Sigma_*=\sqrt{\pi G h_*\Sigma_*}$ is the surface density of the stars and $h_*$ is the disk scale height. Note that this equation is valid only under the condition $\Sigma_*\gg\Sigma_\mathrm{gas}$.\\

\noindent\textit{Star formation in spheroids:} Star formation rates in spheroids are calculated using dynamical times with the same parameters as for the best fit to the more massive galaxies. The timescale for star formation is given by
\begin{equation}
    \tau_*=\epsilon_*^{-1}\tau_\mathrm{dynamical}\bigg(\frac{V}{200\,\mathrm{km/s}}\bigg)^{\alpha_*},
\end{equation}
where $\epsilon_*$ and $\alpha_*$ are input parameters, and $\tau_\mathrm{dynamical}=r/V$ where $r$ and $V$ are the characteristic radius and velocity of the spheroidal component, respectively. This timescale cannot fall below a minimum value of 7.579 Gyrs.\\

\noindent\textit{{Accretion mode onto halos:}} Gas can accrete onto halos in one of two `modes': `cold' and `hot'. In `hot-mode' accretion, all accreted gas is shock heated to the virial temperature of the halo. Although this model describes the process of accretion well for higher mass halos, gas accretion in low mass halos (dwarfs) is never shock heated to the virial temperature \citep{fardal2001,keres2005,keres2009}. Studies such as \cite{keres2005,keres2009} show that `cold-mode' gas accretion dominates low mass galaxies (i.e. $<10^{10.3}\;M_{\odot}$) while `hot-mode' accretion of gas occurs in higher mass systems.  In `cold-mode' accretion, the gas accreted never forms a hydrostatic halo, and so does not need to cool and radiate its thermal energy before flowing into the galaxy. It instead flows into the galaxy on order of the dynamical time. Therefore, we implement `cold-mode' accretion onto low mass halos, `hot-mode' accretion onto high mass halos, and a mixture of both to intermediate mass halos. The transition between two modes is determined by two `shock' parameters. 

According to \cite{BirnboimDekel2003, BensonBower2010}, the cold-mode fraction is defined by
\begin{equation}
    f_\mathrm{cold}=(1+r^{\frac{1}{\delta}})^{-1},
\end{equation}
where $\delta$ is the shock stability transition width, $r=\epsilon_\mathrm{crit}/\epsilon$ and $\epsilon=r_\mathrm{s}\Lambda\rho_sv_s^3$ where $r_\mathrm{s}$ is the accretion shock radius (set to the virial radius), $\Lambda$ is the post-shock cooling function, $\rho_\mathrm{s}$ and $v_\mathrm{s}$ are pre-shock density and velocity (at the virial radius) respectively, and $\epsilon_\mathrm{crit}$ is the accretion shock stability threshold. Here, the pre-shock density is defined by 
\begin{equation}
    \rho_s=\frac{\gamma-1}{\gamma+1}\frac{3}{4\pi}\frac{\Omega_\mathrm{b}}{\Omega_\mathrm{m}}\frac{M}{r_\mathrm{s}^3}\bigg[1+\frac{(\alpha+3)(10+9\pi)}{4}\bigg]^{-1},
\end{equation}
where $M$ is the total halo mass, $\gamma=5/3$ is the adiabatic index of gas, and $\alpha$ is the exponent that corresponds to initial density perturbation \citep{BirnboimDekel2003}.

\subsubsection{Parameters without Astrophysical Priors}\label{subsec:unconstrained}

We next describe the parameters that are unconstrained by the underlying astrophysics of either dwarf galaxies or the Milky Way. Ram pressure and tidal stripping were constrained by comparisons to the observed and simulated \citep{Applebaum2021,Shipp+2022} luminosity functions of the Milky Way satellites. The physics of star formation feedback is constrained to best fit the slope and scatter of the luminosity-metallicity relation.\\

\noindent\textit{{Ram Pressure Stripping:}} We use the model of \cite{font2008} to model ram pressure stripping of hot halo gas in our dwarf galaxies as this method sets a physical radius within the dwarf galaxy halo. The ram pressure stripping radius of \cite{font2008} is a solution to
\begin{equation}
    \alpha  \frac{GM_\mathrm{satellite}(r_\mathrm{rp})\rho_\mathrm{hot,satellite}(r_\mathrm{rp})}{r_\mathrm{rp}}=\mathcal{F}_\mathrm{ram,hot,host},
    \label{eq:ramFsolve}
\end{equation}
where $\mathcal{F}_\mathrm{ram,hot,host}$ is the ram pressure force due to the host halo and $M_\mathrm{satellite}(r)$ is the total mass of the satellite within radius $r$. The ram pressure force due to the hot halo is defined by
\begin{equation}
    \mathcal{F}_\mathrm{ram,hot,host}=\rho_{hot,host}(r)v^2(r).
    \label{eq:ramF}
\end{equation}

\noindent Mass loss rate in disks are computed using the equation
\begin{align}
    \dot{M}_\mathrm{gas,disk}=&\min\left(\frac{\beta_\mathrm{ram}\mathcal{F}_\mathrm{hot,host}}{2\pi G \Sigma_\mathrm{gas}(r_{1/2})\Sigma_\mathrm{total}(r_\mathrm{1/2)},R_\mathrm{max}}\right)\times \\ \nonumber
    &\frac{M_\mathrm{gas,disk}}{\tau_\mathrm{dyn,disk}},
\end{align}
where $\beta_\mathrm{ram}$ is the ram pressure stripping efficiency which scales the mass loss in the disk, $\Sigma_{gas}(r)$ is the gas surface density in the disk, $\Sigma_{total}(r)$ is the total surface density in the disk, $r_{1/2}$ is the disk half mass radius, $M_{gas,disk}$ is the total gas mass in the disk, $\tau_{dyn,disk}$ is the dynamical time in the disk, $R_{max}$ determines the maximum rate of gas mass lost, and G is the gravitational constant.

In spheroids, the rate of gas mass loss is calculated using
\begin{equation}
    \dot{M}_\mathrm{gas,sph}=-\max(\alpha,R_\mathrm{max})M_\mathrm{gas}/\tau_\mathrm{sph},
\end{equation}
where $\alpha=\beta_\mathrm{ram}\mathcal{F}_\mathrm{hot,host}/F_{gravity}$, $M_\mathrm{gas}$ is the mass of gas in spheroid and $\tau_\mathrm{sph}$ is the dynamical time of the spheroid. The gravitational restoring force at half mass radius is given by 
\[
F_{gravity}=\frac{4}{3}\rho_\mathrm{gas}(r_{1/2})\frac{GM_\mathrm{total}(r_{1/2})}{r_{1/2}}.
\]


\noindent\textit{{Tidal Stripping:}} There is evidence for tidal stripping in dwarf satellites embedded in the scatter of the halo-stellar mass relation \citep{jackson2021} and the presence of tidal streams and debris \citep{BullockJohnston2005}. Previous studies have shown that more dark matter must be stripped in order for stripping of stars to occur in galaxies with smaller disks \citep{penarrubia2008,Smith+2013}. Simulations suggest that stars in dwarf spheroids are only stripped after $80-90\%$ of the dark matter is stripped \citep{Smith+2013}. In addition, galaxies that lose 80\% of dark matter mass lose about 10\% of their stellar mass \citep{rory+2016}. As such, tidal stripping of dark matter precedes tidal stripping of stars. 

We approximate stellar mass and ISM gas loss via tidal stripping treatment using the `simple' model in Galacticus. This model assumes the stellar mass loss rate scales with the ratio of tidal force to restoring force in a galaxy at half mass radius, and is inversely proportional to the dynamical timescale
\begin{equation}
    {\dot M_*}=\beta_\mathrm{tidal}\frac{F_\mathrm{tidal}}{F_\mathrm{res}}\frac{1}{T_\mathrm{dyn}}M_*,
\end{equation}
where $\beta_\mathrm{tidal}$ is the strength of tidal stripping of ISM and stars, $F_\mathrm{tidal}$ is the tidal force, $F_\mathrm{res}$ is the restoring force, $T_\mathrm{dyn}$ is the dynamical time of stars, and $M_*$ is the stellar mass. Note that this model only captures the effects of tidal stripping on the total mass and ignores the effects on the shape of the galaxy's density profile.\\

\noindent\textit{{Stellar Feedback:}} We next determine the parameterization of stellar feedback which best produces the observed luminosity-metallicity relation. Stellar feedback from the disk and spheroid components are treated separately, but with the same model, parameterized by a characteristic velocity and exponent. The characteristic velocity defines the scale at which supernovae feedback results in a mass-loading factor (the ratio of the outflow rate to the star formation rate) is one.
The outflow rate is then given by
\begin{equation}
\dot M_\mathrm{outflow}=\bigg(\frac{v_\mathrm{outflow}}{v}\bigg)^{\alpha_\mathrm{outflow}}\frac{\dot E_*}{E_\mathrm{canonical*}},
\end{equation}
where $v_\mathrm{outflow}$ (the disk/spheroid characteristic velocity) and $\alpha_\mathrm{outflow}$ (the disk/spheroid exponent) are tunable parameters, $\dot{E}_*$ is the rate of energy input from stellar populations, and $E_\mathrm{canonical*}$ is the total energy input by a canonical stellar population normalized to $1\;M_{\odot}$ after infinite time.




\section{Results}\label{sec:results}

We initially explore whether there is a set of input parameters for which running Galacticus on a high resolution N-body merger trees which can reproduce the luminosities and metallicities of the Milky Way dwarfs. 

\subsection{Luminosity Function}
 
We begin our exploration of the best fit Galacticus parameters by determining the combination of $v_\mathrm{cooling}$ and $v_\mathrm{filter}$ which best reproduce the observed luminosity function of the Milky Way dwarfs and the simulated luminosity functions from the Mint Justice League \citep{Applebaum2021} and FIRE II mock observations \citep{Shipp+2022}. For MINT Justice League, we use Sandra ($2.4\times 10^{12}\;M_{\odot}$) and Elena ($7.5\times 10^{11}\;M_{\odot}$) since they are the only simulations run at MINT resolution. They also have virial masses closest to our Milky Way analog ($1.8\times10^{12}\;M_{\odot}$ or $1.2\times10^{12}\;M_{\odot}/h$). We compare our models to 3 Milky Way analogs (m12f, m12m, m12i) of FIRE II hydrodynamic simulations. Masses of m12f, m12m, m12i are $1.7\times10^{12}\;M_{\odot},1.6\times10^{12}\;M_{\odot},\;1.2\times10^{12}\;M_{\odot}$ respectively.

While the luminosity functions of the hydrodynamical simulations may undercount the number of ultra-faint dwarfs due to over merging \citep{graus2019}, they do not have the completeness issues of the observation sample \citep{drlica-wagner2020}.

\begin{figure*}[htb!]
\centering
\includegraphics[clip=true,trim={4cm 0cm 0 4cm},width=19cm]{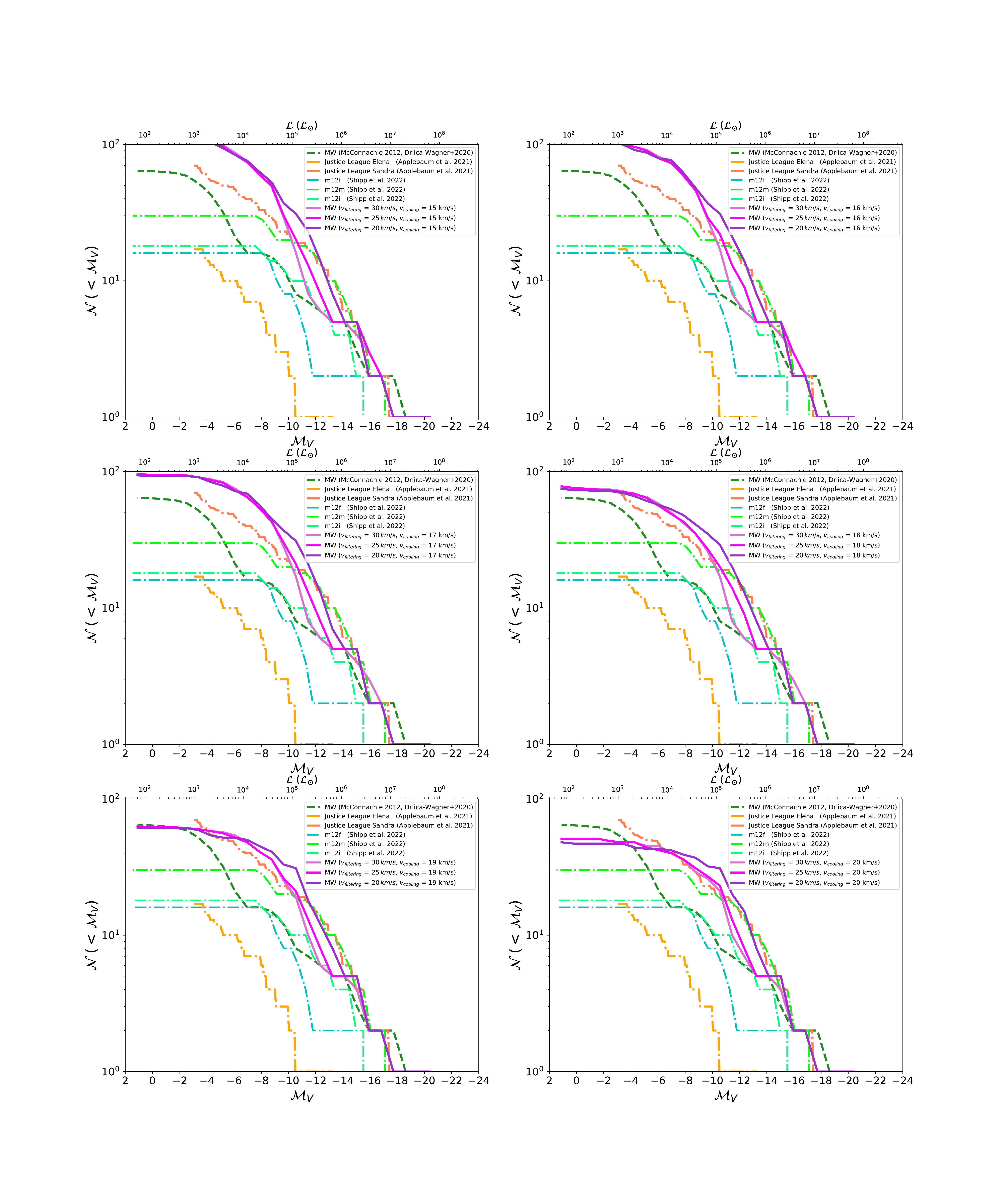}
\vspace{-2cm}
\caption{Cumulative luminosity function of the Milky Way dwarf satellite galaxies. $M_V$ denotes the absolute V band magnitude and $\mathcal{N}$ denotes the cumulative number of galaxies fainter than $M_V$. The dark green dashed line shows the observed data from \cite{mcconnachie2012,drlica-wagner2020}. Each figure corresponds to Galacticus runs with cooling rate cutoff velocities from $15-20\,km\,s^{-1}$. These predicted luminosity functions correspond to $v_\mathrm{filter}\,=\,20,25,30\,km/s$ are then compared to Justice League hydro simulations (shown in orange and coral), and mock observations of FIRE II hydro simulations (in shades of green) along with observations.}
\label{fig:Mv_cooling_vF}
\end{figure*}

Figure~\ref{fig:Mv_cooling_vF} shows the luminosity function of the satellites in our Milky Way analog modeled by Galacticus with $v_\mathrm{cooling}~=~15~-~20$~km~s$^{-1}$ and $v_\mathrm{filter}~=~20~-~30$~km~s$^{-1}$. It shows the effect of our choices of $v_\mathrm{cooling}$ and $v_\mathrm{filter}$, for $z_\mathrm{reion}\,=\,9$. The filtering velocity is only allowed to range from $20$~km~s$^{-1}$ to $30$~km~s$^{-1}$ \citep{gnedin2006,Bovill&ricotti2011}. The choice of the range of $v_\mathrm{cooling}$ and $v_\mathrm{filter}$ approximates the known physics which suppresses gas accretion and cooling in low mass halos. In this work, we hold the reionization redshift of the Milky Way constant. 

To determine the combinations of $v_\mathrm{cooling}$ and $v_\mathrm{filter}$ which produce the best agreement with the known Milky Way satellite population, we compare our models to the observed luminosity function \citep{mcconnachie2012,drlica-wagner2020} and the simulated luminosity function from the two halos in the MINT Justice League \citep{Applebaum2021} simulations. The latter minimizes the complications due to the incompleteness of the sample of Milky Way satellites, especially at $M_V~>~-10$ \citep{willman+2004}. Note that we use the updated version of \cite{mcconnachie2012} as of January 2021. \cite{drlica-wagner2020} attempts to correct for the survey incompleteness to find the total number of dwarfs in DES and  Pan-STARRS1(PS1) surveys. We find that $v_\mathrm{cooling}$ plays a critical role in producing the correct number of dwarf galaxies fainter than $M_\mathrm{V}\geq-8$, while $v_\mathrm{filter}$ primarily affects brighter dwarfs. Our model produces the best fit to the luminosity function of the Milky Way satellites ($M_\mathrm{V}~<~-6$) and Sandra \citep{Applebaum2021} for $v_\mathrm{cooling}=18-19$~km~s$^{-1}$ with $v_\mathrm{filter}~=~25$~km~s$^{-1}$. 

For our best fit model, we match the number of galaxies fainter than $M_V\,=\,-12$ (Figure~\ref{fig:Mv_cooling_vF} bottom left panel). However, the number of brighter satellites are under-predicted in comparison to observations and hydrodynamic simulations (middle line shown in pink). As the number of bright satellites around a Milky Way mass host is low, this is may simply be due to small number statistics. 
    

Notice that there is more than one set of parameters for $v_\mathrm{cooling}$ and $v_\mathrm{filter}$ which will produce a reasonable fit to the luminosity function of the observed Milky Way satellites and the MINT Justice League. Specifically, our fit is not improved markedly for $v_\mathrm{filter}=25$--30~km/s and $v_\mathrm{cooling}\sim18$--20~km/s. In this work, we choose our best fit value for $v_\mathrm{cooling}$ to approximate the atomic cooling threshold during the epoch of reionization. Our `best fit' $v_\mathrm{filter}$ is chosen to be the average of the values used in \cite{RicottiGnedin:05} and \cite{Bovill&ricotti2011}.

\begin{figure}[h!]
    \centering
    \includegraphics[clip=true,trim={1cm 0cm 0 1cm},width=8.5cm]{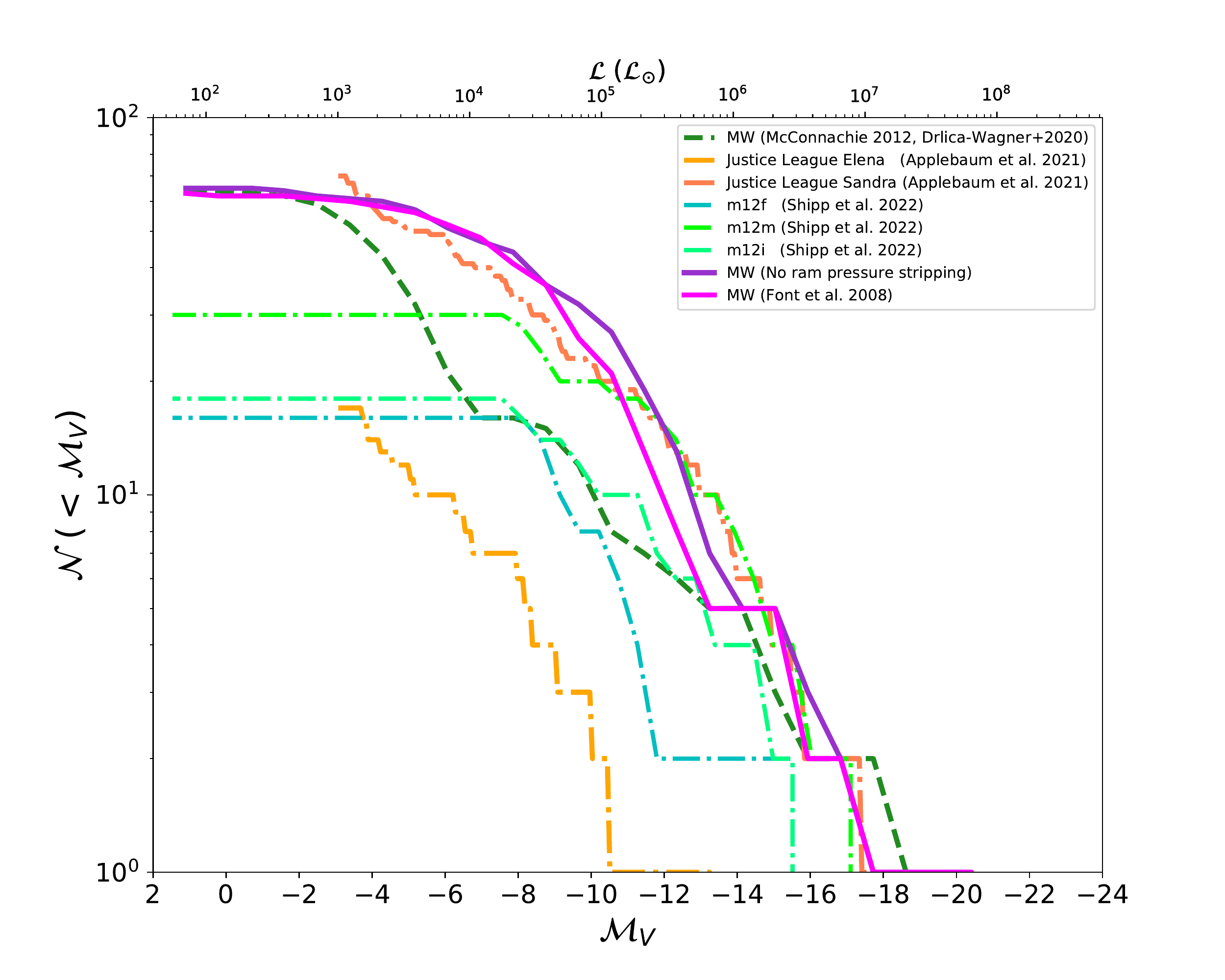}
    \caption{Cumulative luminosity function of the Milky Way satellites computed with ram pressure stripping methods \citep[][pink]{font2008} and no ram pressure stripping (purple). Other colors are the same as in Fig~\ref{fig:Mv_cooling_vF}.}
    \label{fig:ram_Font_vR}
\end{figure}

In Figure \ref{fig:ram_Font_vR} we compare the cumulative luminosity function for the Milky Way satellites computed with the \cite{font2008} ram pressure stripping model (pink), to that computed in a model with no ram pressure stripping (purple).

We now look at the effect of ram pressure stripping for our best fit cooling and filtering velocities. We vary the efficiency of the ram pressure stripping through its full range from 0 to 1. However, the effect of $\beta_{ram}=0.01$ seem to be same as $\beta_{ram}=1.00$ i.e. the efficiency at which gas is stripped upon infall does not have a major effect on the luminosity except in the more massive dwarfs. Note that, with the exception of some minor differences at high luminosity ($M_V~<~-14$), changing the efficiency of the ram pressure stripping does not significantly affect the luminosities of our modeled galaxies. This is expected, as only the most massive Milky Way dwarfs formed significant amounts of stars after their infall into the Milky Way halo \citep{Rocha+2012}.

\begin{figure}[h!]
\centering
\includegraphics[clip=true,trim={1cm 0cm 0cm 1cm},width=8.5cm]{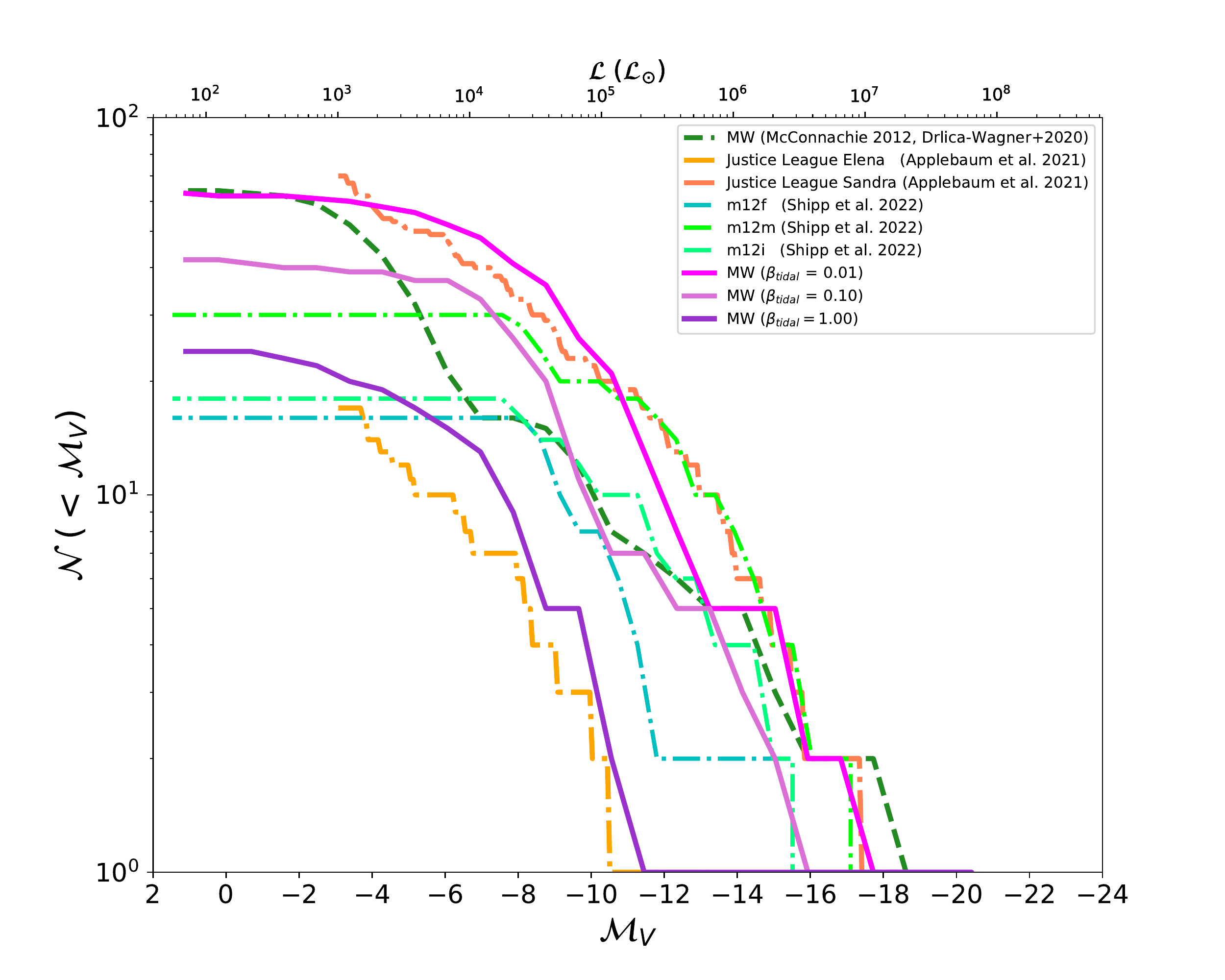}
\caption{Luminosity function of Milky Way dwarfs for varying tidal stripping efficiencies. Three colors violet, fuchsia, and purple indicate tidal stripping efficiencies for stars and ISM gas ($\beta_\mathrm{tidal}=$ 0.01, 0.1, and 1 respectively). Lower efficiency is in agreement with observations and results of MINT resolution Justice League, and FIRE-II hydrodynamic simulations (colors are the same as in Fig~\ref{fig:Mv_cooling_vF}).}
\label{fig:tidal_Mv}
\end{figure}
We now move onto tidal stripping using the `simple' model in Galacticus. Since, in the N-body simulation, there is already stripping of the dark matter halos, we do not implement any additional stripping of the dark matter.

The strength of tidal stripping of ISM gas and stars $\beta_\mathrm{tidal}$ can be varied from 0 to 1. Unlike ram pressure stripping, which was insensitive to our choice of $\beta_{ram}$, Figure \ref{fig:tidal_Mv} shows the effect on our luminosity function when the efficiency of tidal stripping is varied. We reproduce the observed and simulated (hydrodynamic) luminosity functions with $\beta_\mathrm{tidal}\sim 0.01$. We find a strong and direct, inverse relationship between the efficiency of the tidal stripping and the luminosity function of the Milky Way satellites. 

Note that the tidal force in the model is calculated at the pericenter of satellite's orbit. Therefore the actual tidal force will likely be lower than our estimate. This means that $\beta_\mathrm{tidal}~<<~1$ is reasonable. In addition, models suggests the majority of the dark matter must be stripped before the stars \citep{penarrubia2008,rory+2016}. 
Since all the dwarf galaxies in our simulation exist in {\it{intact}} dark matter halos, this is in line with expectations from \cite{penarrubia2008} that $>90\%$ of the dark matter halo needs to be stripped before the stars are significantly affected. Our model currently includes only a few halos which have been stripped to this level, thus low efficiency of tidal stripping used here is in agreement with previous work.\\

\subsection{Luminosity Metallicity Relation}

\begin{figure}[h!]
\centering
\includegraphics[clip=true,trim={0cm 0cm 0 1cm},width=8.5cm]{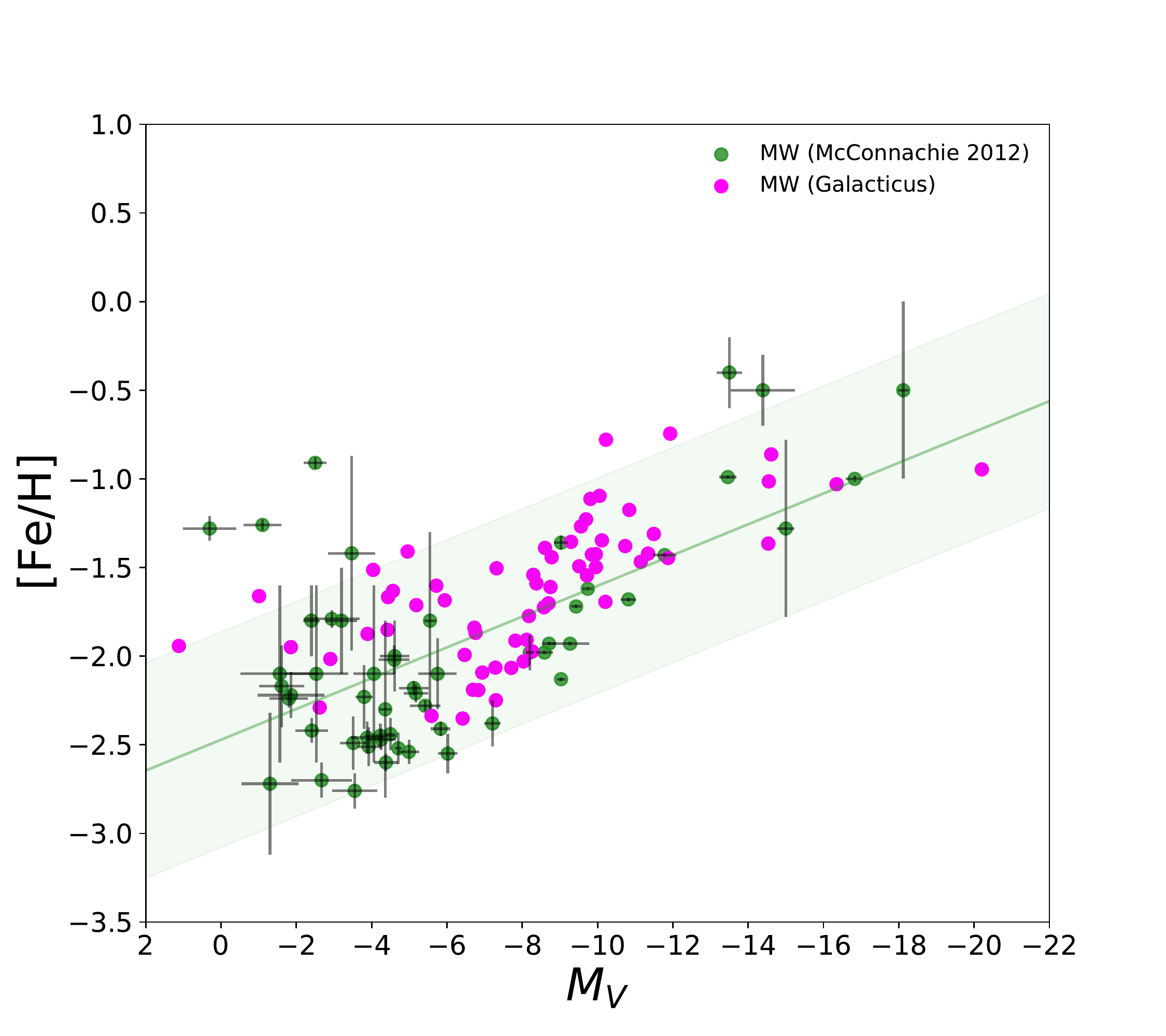}
\caption{Iron abundance of the dwarf satellite galaxies as a function of absolute V-band magnitude. Observed data from \cite{mcconnachie2012} are green and dwarfs modeled with Galacticus are pink. Exisiting stellar feedback recipies in Galacticus has been calibrated to reproduce the luminosity-metallicity relation.}
\label{fig:FeH}
\end{figure}

\begin{figure*}[!h]
    \centering
    \includegraphics[width=16cm]{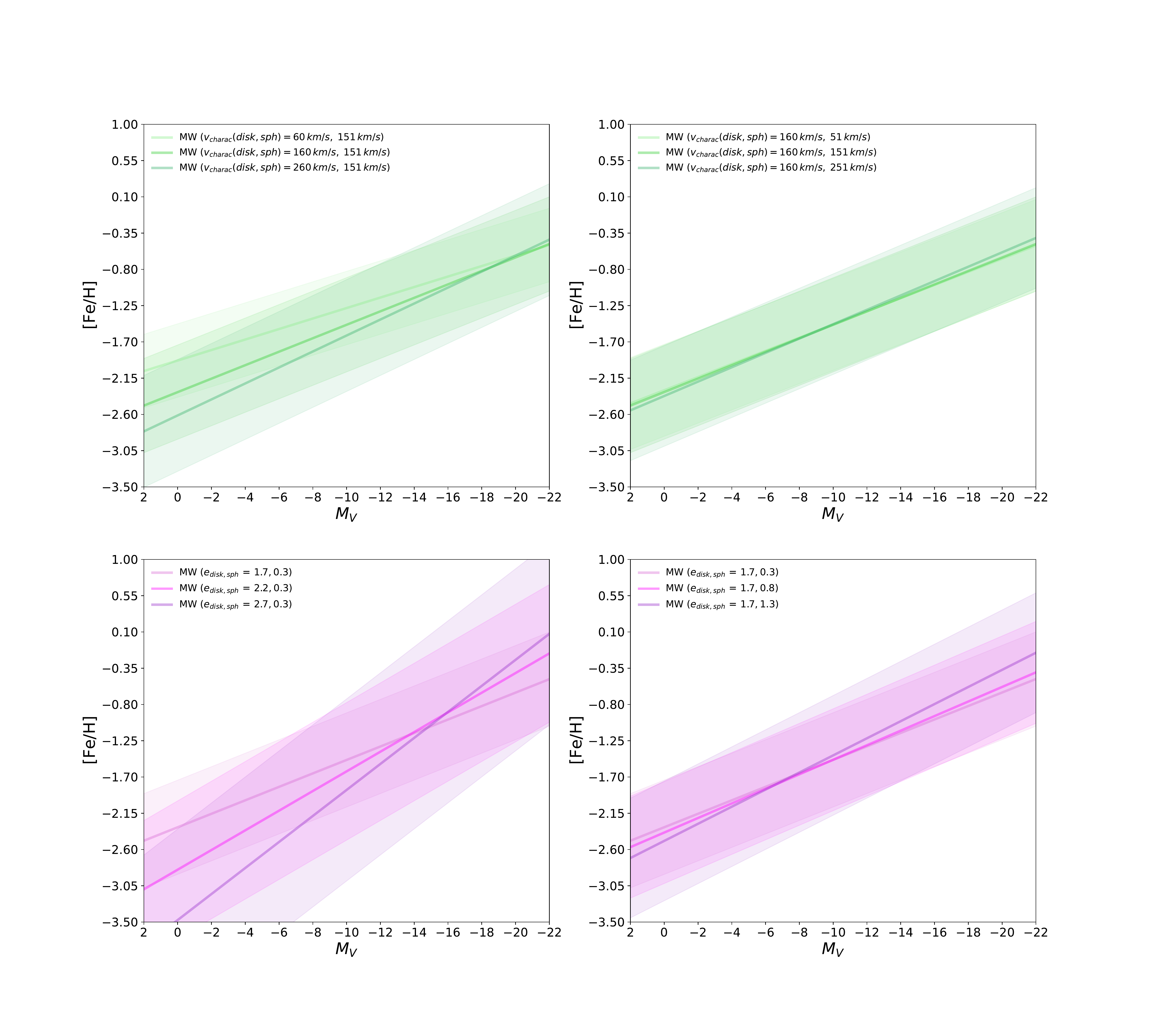}
    \caption{Modeled luminosity-metallicity relations for various characteristic velocities and exponents of stellar outflows. Top left and right figures show the effect of the characteristic velocity for the disk and spheroid components respectively. Bottom left and right figures show the effect of exponents on disk and spheroid components. Note that this relation is sensitive to both characteristic velocity (normalization) and exponents of the disk component (slope).}
    \label{fig:v_charac}
\end{figure*}

We next determine the combination of stellar feedback parameters which best reproduces the slope of the observed luminosity-metallicity relation \citep{mcconnachie2012}. We tune our model for $v_\mathrm{characteristic,disk}=60,160,260\;km/s$ for the disk and $v_\mathrm{characteristic,sph}=51,151,251\;km/s$ for the spheroidal component. We find that by tuning the existing stellar feedback recipes in Galacticus we can reproduce both the trend and scatter in the observed luminosity-metallicity relation (Figure~\ref{fig:FeH}). Critically, metallicities of the modeled dwarfs match well with observations down to the ultra-faint dwarfs. The two exponents, $\alpha_\mathrm{outflow,disk}$ and $\alpha_\mathrm{outflow,spheroid}$,and  the circular velocity ($V$) determine the scaling of the outflow rate of the corresponding disk/spheroid measured at the scale radius of that component. The characteristic velocity determines normalization of the luminosity-metallicity relation, and exponent of the disk, tuneS the slope (Figure \ref{fig:v_charac}). Higher exponents correspond to steeper slopes and vice versa. In particular, low mass dwarf galaxies are sensitive to exponents controlling their supernova-driven outflows. The closest match to the slope to observed luminosity-metallicity relation is obtained for exponents $\alpha_\mathrm{outflow,disk}=1.7$ and $\alpha_\mathrm{outflow,spheroid}=0.3$.

As seen in Figure~\ref{fig:v_charac}, while the exponent for the spheroid only marginally affects the slope of the luminosity-metallicity relation, the effect of tuning stellar feedback in the disk component is far greater. We find stellar feedback outflows to be a significant component for tuning the luminosity-metallicity relation. This agrees with \cite{lu2015} which demonstrated that metallicity of galaxies provides a constraint on the maximum outflow velocity ($\sim 141 \;km/s$).

\subsection{Properties of the Milky Way Dwarfs}

We have determined a set of parameters for Galacticus which reproduce the observed luminosities and metallicities of the Milky Way dwarfs. In this section, we determine if these parameters can reasonably reproduce other properties of the Milky Way dwarfs. Unlike the luminosity function and luminosity-metallicity relation discussed above, we have {\it{not}} tuned Galacticus to reproduce any of the dwarf galaxy properties below. All the observational data in this section comes from the updated table as of Jan 2021 originally published in \cite{mcconnachie2012}.

\subsubsection{Half Light Radii}

\begin{figure}[h!]
    \centering
    \includegraphics[width=8cm]{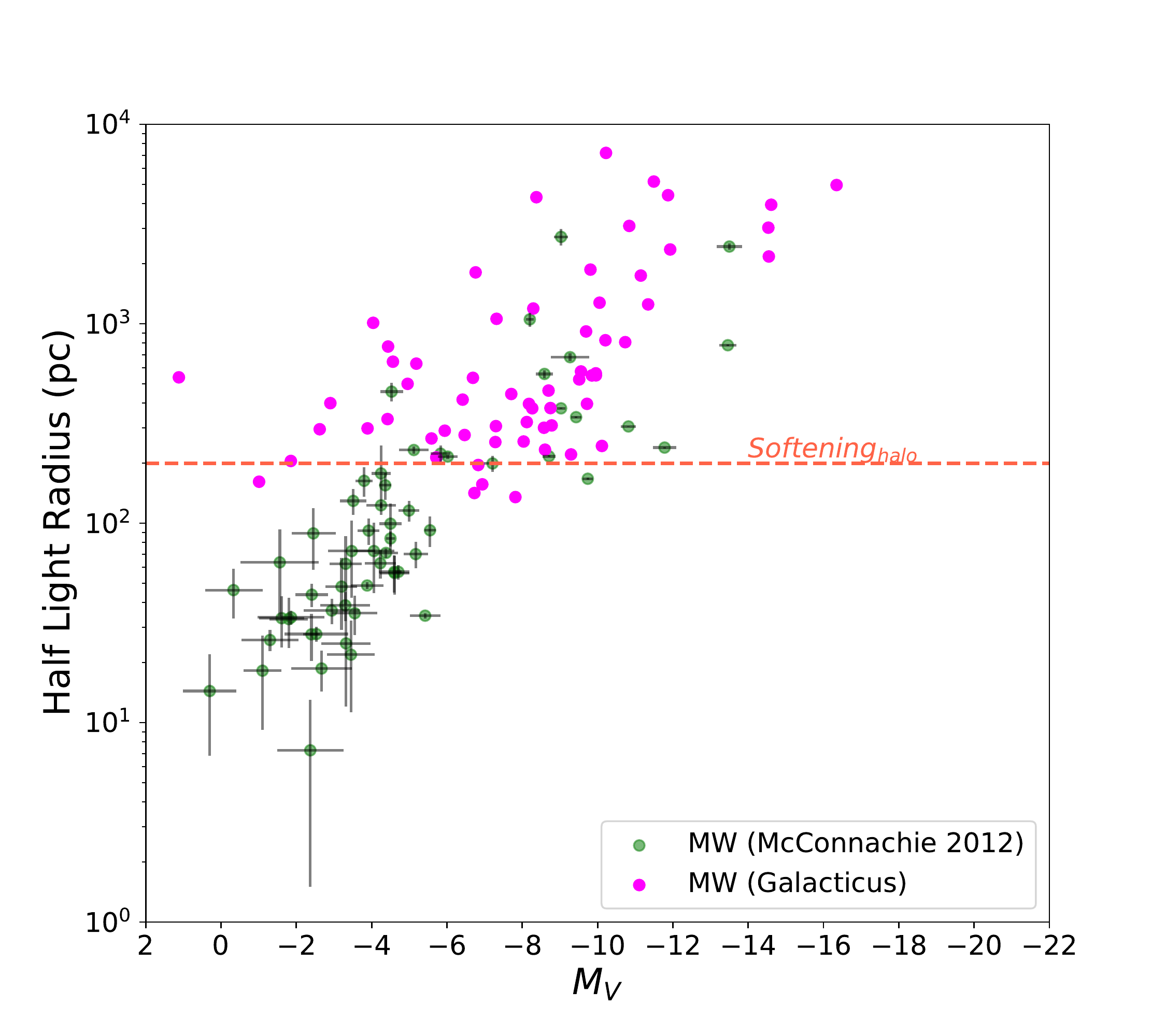}
    \caption{Half light radii of the dwarf satellite galaxies as a function of absolute V band magnitude. Observed data from \cite{mcconnachie2012} are colored in green, and simulated data are colored in pink. Orange dashed line shows the softening of the halo at $200\,pc$ in the N-body simulation.}
    \label{fig:hlr}
\end{figure}

As seen in Figure~\ref{fig:hlr}, we are able to match the observed half-light radii for the Milky Way satellites down to $M_\mathrm{V} \leq -6$. However, our modeled dwarfs have larger half-light radii for fainter, smaller dwarfs, and our modeled half-light radii do not reach below $~200\;pc$. This `floor' in our half-light radii roughly corresponds to the physical softening of our simulations (orange line in Figure~\ref{fig:hlr}).

\begin{figure}[h!]
\centering
    \includegraphics[clip=true,trim={0cm 0cm 0cm 1cm},width=9cm]{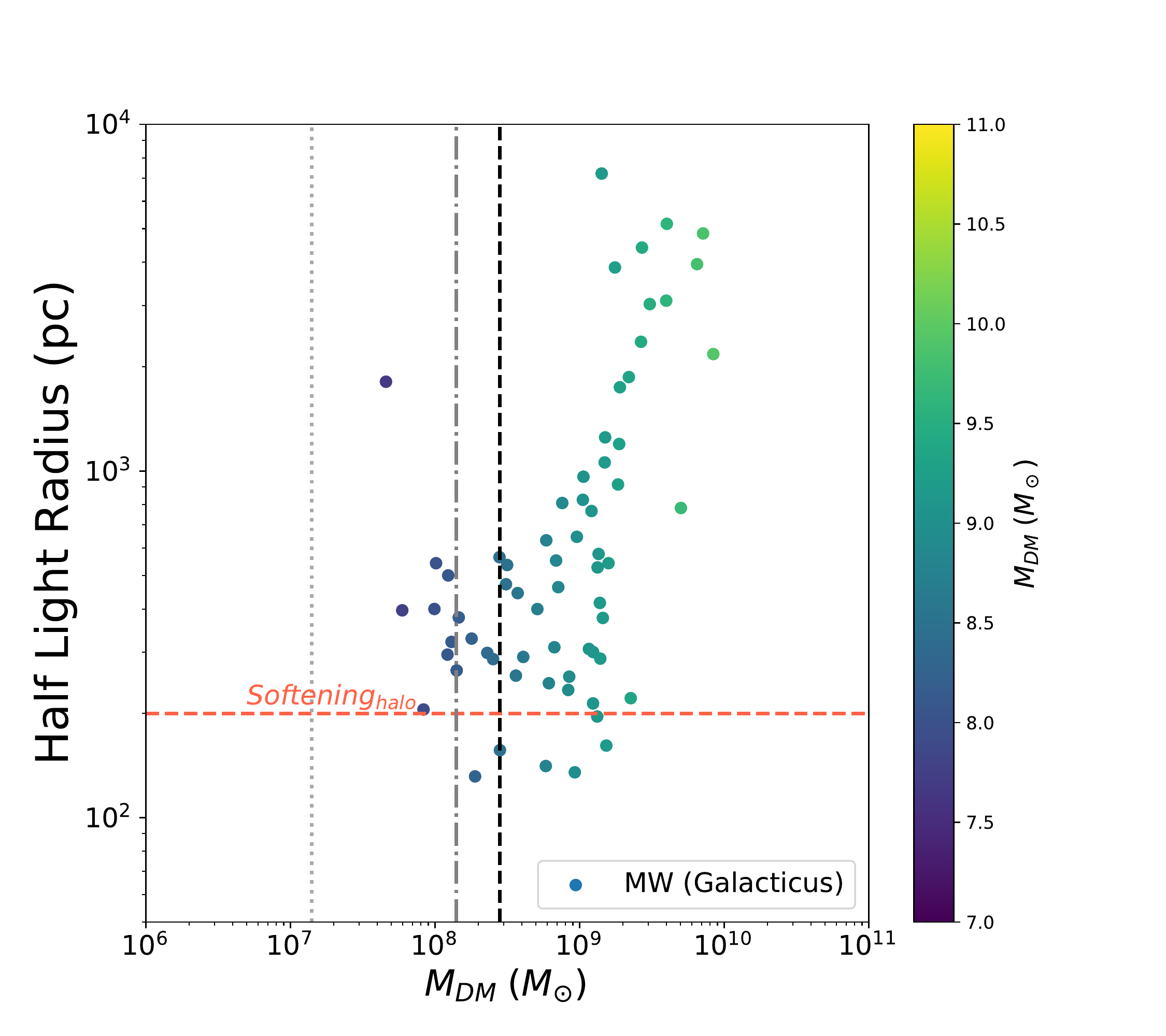}
    \caption{Half-light radii of the dwarf satellite galaxies as a function of dark matter halo mass. Here we show the dark matter mass of halos with 100 (light grey dotted), 1000 (dotted dashed line in dark grey), and 2000 (dashed line in black) particles. The orange dashed line shows the softening of the halo at $200\,pc$ in the N-body simulation.}
    \label{fig:hlr_DM}
\end{figure}

In order to investigate this, we look at the half light radii as a function of dark matter halo mass (Figure~\ref{fig:hlr_DM}). The vertical lines in Figure~\ref{fig:hlr_DM} show the dark matter halo masses for various numbers of particles per halo. Note that halos whose half-light radii are below the `floor' corresponding to the physical softening of our simulation all have $>~1000$ particles. As Galacticus calculates the $r_\mathrm{hl}$ of the halos by allowing the disk and spheroidal components to evolve within the gravitational potential of a dark matter profile, the determination of $r_\mathrm{hl}$ relies on a robust determination of the dark matter profile. The underlying NFW profile is set from scale radii of the simulation, where concentrations are calculated using the model by \citep{Gao+2008}. The equilibrium radii for the disk and the spheroid components are described by the NFW profile, and half-light radii are calculated in $SDSS_g$ luminosity band. While the global properties of halos with $N~<~1000$ particles are relatively certain \citep{2010ApJ...711.1198T,2017MNRAS.467.3454B}, the details of their dark matter profiles are not robust. For example, \cite{2021MNRAS.500.3309M} show that convergence in measurements of half-mass radii of halos from N-body simulations requires $>~4000$ particles.  


As the low mass halos which host the faintest dwarfs in our model have $N~<~500$ particles, the uncertainties in the determination of their dark matter profile coupled with the physical gravitational softening used in the simulation produces a `floor' of $\sim 200\;pc$. Similar effects of resolution are seen in the half-light radii of MINT Justice League simulations by \cite{Applebaum2021}.

\subsubsection{Velocity Dispersion}

\begin{figure}[h!]
    \centering
    \includegraphics[width=8cm]{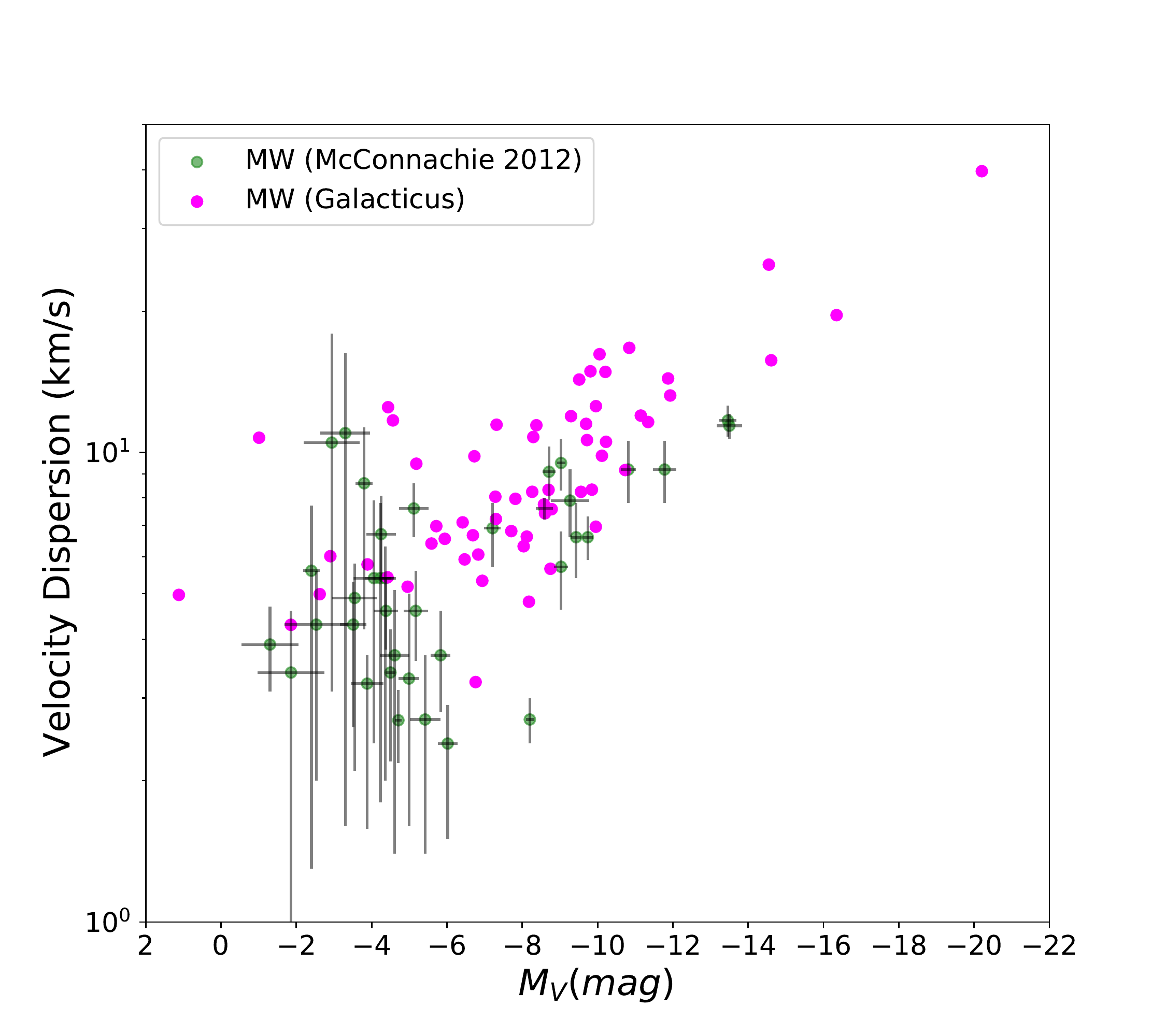}
    \caption{Velocity dispersion of the modeled and observed dwarf satellites as a function of absolute V band magnitude. Observed data from \cite{mcconnachie2012} are colored in green, and simulated data are colored in pink. Our model agree well with observations without additional tuning.}
    \label{fig:Vdis}
\end{figure}

We next look at the velocity dispersions of our modeled dwarfs at half-stellar mass radii compared to observations of \cite{mcconnachie2012} (updated as of Jan 2021). As seen in Figure~\ref{fig:Vdis}, the stellar velocity dispersions of our predicted dwarfs agree well with observations. However, note that velocity dispersions of galaxies below $M_V\sim-8$ may be affected by the the floor in half-light radii discussed above.

\subsubsection{Mass to Light Ratios}

\begin{figure}[!h]
    \centering
    \includegraphics[width=8cm]{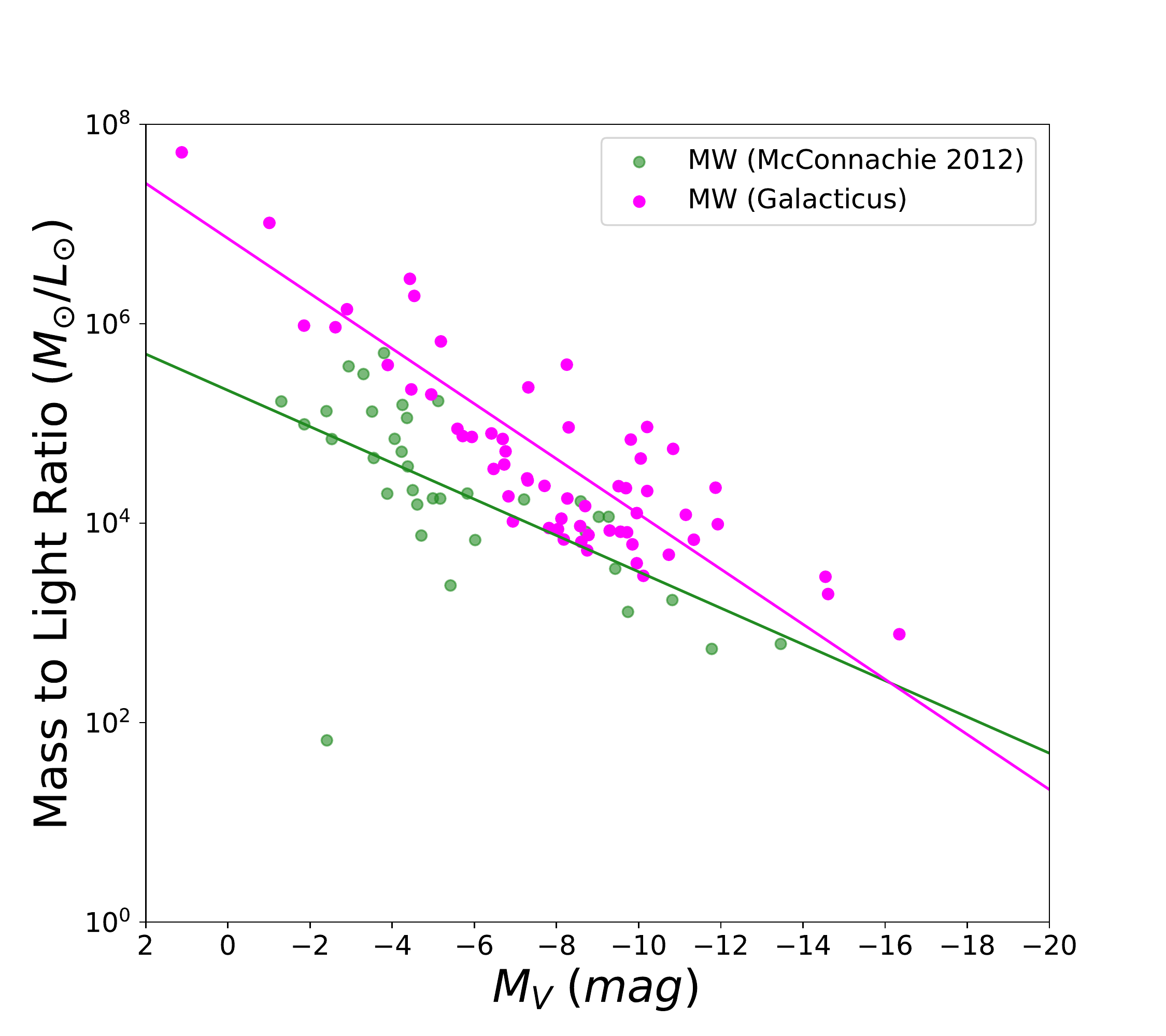}
    \caption{Mass to light ratios of the Milky Way satellites as a function of the velocity dispersion along the line of sight. Mass to light is calculated using the half mass with velocity dispersion and half light radii as described in \cite{wolf2010}. We compare the modeled dwarfs (pink) to observed data from \cite{mcconnachie2012} (green). Green and pink lines show the linear regression lines for the observed and modeled dwarfs respectively. Mass to light ratios of the Milky Way satellites are in agreement down to the ultra faints, though our mass to light ratios are a bit higher than observed values.}
    \label{fig:mass to light}
\end{figure}

Given that we are able to reasonably reproduce the half-light radii and velocity dispersion of the Milky Way dwarfs, we can estimate the mass-to-light ratios of the modeled Milky Way satellites using the method given in \cite{wolf2010}. Our modeled mass-to-light ratios are in good agreement with values derived from observations (Figure \ref{fig:mass to light}). Critically, we are able to produce the dark matter domination of the ultra-faints dwarfs \citep{Simon2019}. We quantify the offset in observational and simulation data by using two regression lines (see Figure \ref{fig:mass to light}). We find the root mean square error of 0.84 in log scale.

\subsection{Star Formation Histories}

We have shown that by constraining Galacticus to reproduce the luminosities and metallicities of both the classical and ultra-faint dwarfs, we are able to successfully reproduce a wide range of observed Milky Way dwarf properties at $z~=~0$. As a final test, we determine whether we are able to reproduce star formation histories which match those derived from observations \citep{weisz2014,weisz2015}.

\begin{figure*}[h!]
    \centering
    \includegraphics[clip=true,trim={4cm 0 0 0},width=20cm]{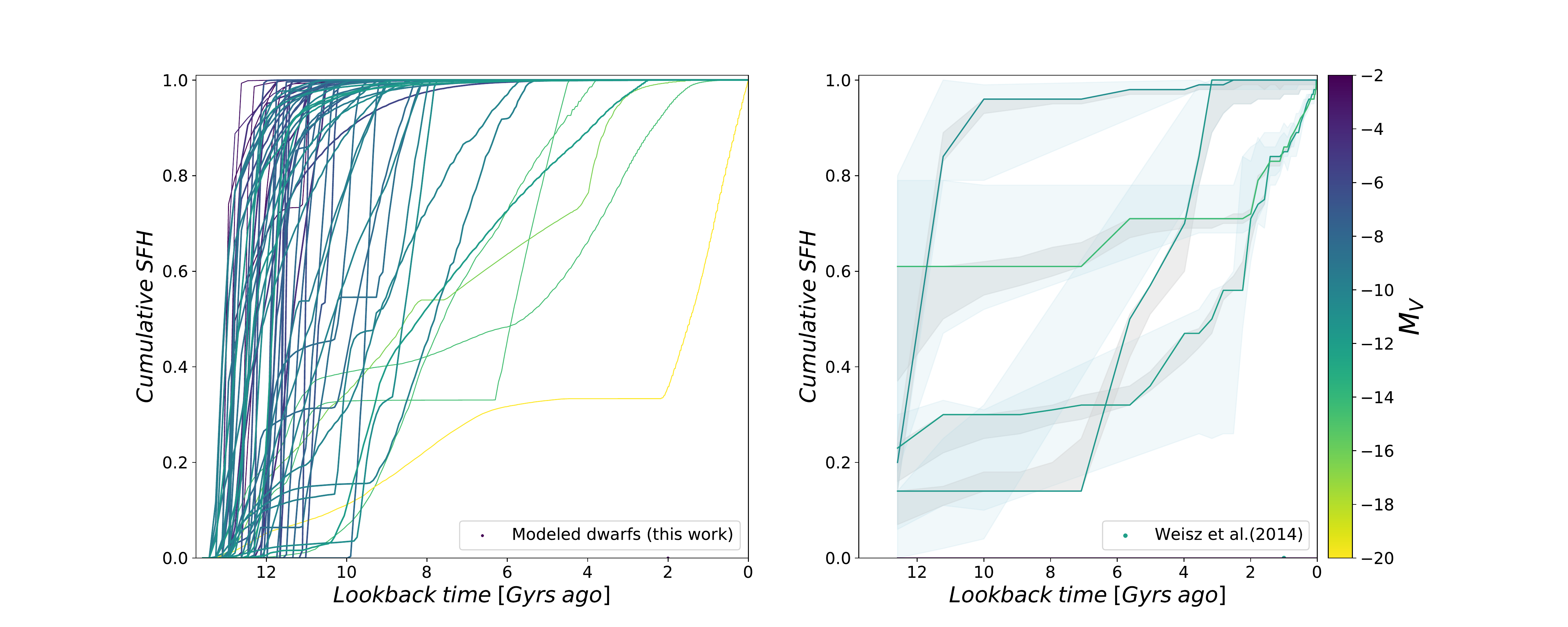}
    \caption{Cumulative star formation histories (SFHs) of the Milky Way satellites colored by absolute V band magnitude. The left panel shows the SFHs modeled with Galacticus and the right panel shows observed SFHs for Leo A, Sagittarius, Sex A, and Sculptor by \cite{weisz2014}. The shaded regions in the right panel shows the region between the 16th and 84th percentiles in SFH uncertainties (grey for random uncertainty and blue for total uncertainty). Note, that \cite{weisz2014} use isochrones older than the age of the universe, and sets the cumulative SFHs to 0 at $log(t)=10.15\;Gyrs$. We have not made any correction to account for this in the modeled SFHs. Our cumulative SFHs are somewhat consistent with these results and the SFHs of the ultra-faints \citep{brown2014}.}
    \label{fig:SFH}
\end{figure*}

We begin by looking at the cumulative star formation histories of the Milky Way dwarfs grouped by $z~=~0$ absolute V band magnitude (Figure \ref{fig:SFH}). Each curve is color-coded by absolute V band magnitude of a particular halo at $z\,=\,0$.  

As expected \citep{brown2014,sacchi2021}, fainter dwarfs ($M_\mathrm{V}\geq-6$) accumulate the majority of their current stellar mass more than $11\pm 1$ Gyrs ago. In contrast, the more luminous model dwarfs at $z~=~0$ form their stars over longer periods of time, including some systems which are still star forming today. We note that some of the modeled cumulative SFHs plateau around 0.1, 0.3, and 0.6 which is similar to observed SFHs of \cite{weisz2014} shown in Figure \ref{fig:SFH}. Cumulative star formation histories derived from Galacticus are consistent with the results of MINT Justice League simulations (Figure~11 of \cite{Applebaum2021}).

The faintest modeled dwarfs all have their star formation cut off at about the same time. This is expected as the faintest observed Milky Way satellites are the fossils of the first galaxies \citep{Bovill&ricotti2011,brown2012}. The larger range the lookback time of the truncation of star formation for the more luminous dwarfs is consistent with their star formation being shut off upon accretion into the Milky Way halo.

\begin{figure*}[h!]
    \centering
    \includegraphics[width=18cm]{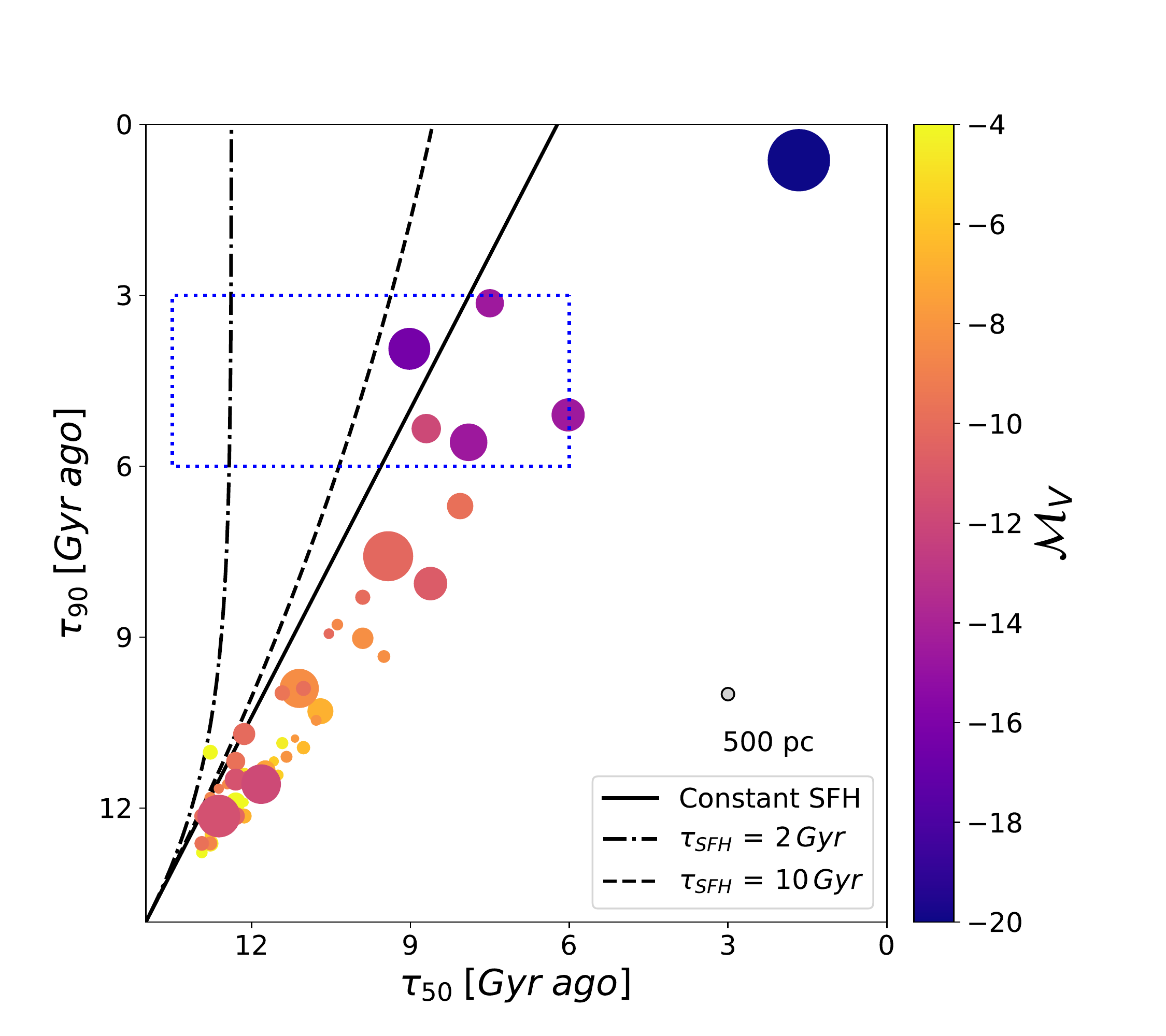}
    \caption{Look back time at which $90\%$ of the stellar mass formed ($\tau_{90}$) versus the look back time at which $50\%$ of the stellar mass formed ($\tau_{50}$). Each point is colored by its absolute V-band magnitude at $z\,=\,0$ and sized relative to their half light radii in pc. The grey dot shows a point with half light radius of 500~pc. The solid line shows constant star formation history. The two dashed lines correspond to exponentially declining SFH (e.g. $SFH(t)=t_0\cdot e^{-t/2\,Gyrs}$ and $SFH(t)=t_0\cdot e^{-t/10\,Gyrs}$ respectively, where $t_0$ is a constant. Compare this plot to the Figure 3 in \cite{weisz2019}. \cite{weisz2019} uses the rectangle shown in blue to show the region within with there are no Milky Way satellites, in contrast to the M31 system. }
    \label{fig:t50_t90}
\end{figure*}

We next look at quenching times of these dwarf galaxies, specifically, the time for a galaxy to gain 90\% its current stellar mass ($\tau_{90}$) and for a galaxy to gain 50\% its current stellar mass ($\tau_{50}$). We reproduce the $\tau_{90}$ versus $\tau_{50}$ plot from Figure 3 of \cite{weisz2019} to compare the overall distribution of star formation histories of the modeled versus observed dwarfs. Interestingly, we are unable to reproduce the lack of galaxies inside the blue dotted rectangle in Figure \ref{fig:t50_t90}, which is a feature \cite{weisz2019} identifies in the Milky Way dwarfs, however no such feature if seen for the M31 dwarfs, in agreement with Figure \ref{fig:t50_t90}.

\begin{figure}[h!]
    \centering
    \includegraphics[clip,trim={7cm 0cm 6cm 0cm},width=8cm]{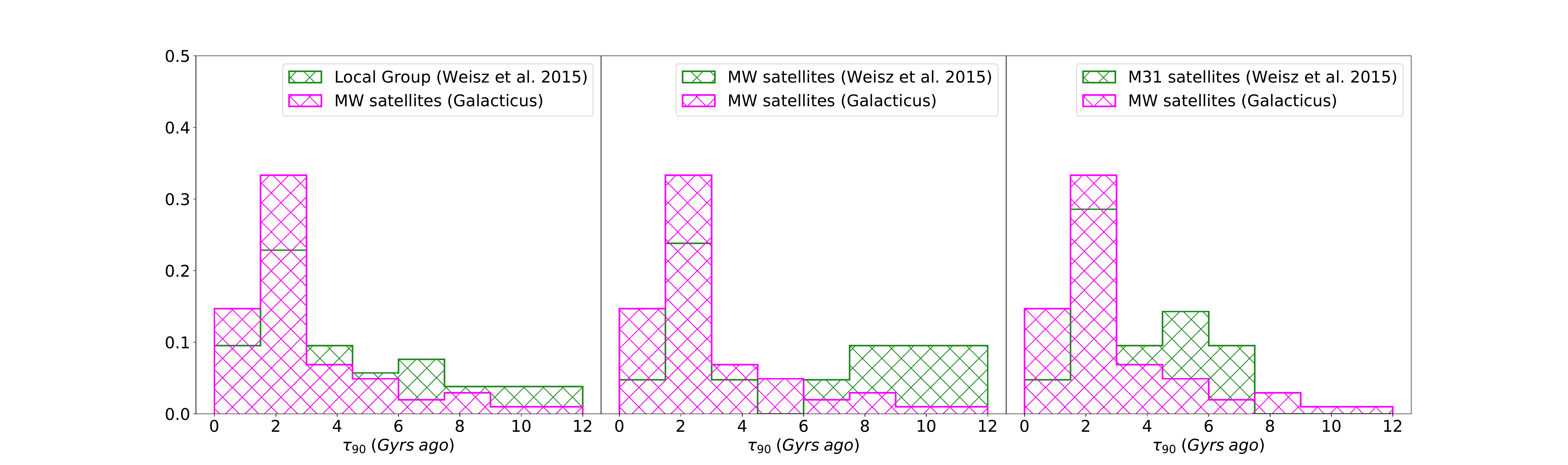}
    \caption{Normalized distribution of $\tau_{90}$ in Gyrs. Two curves show the predicted values from observations of \cite{weisz2015} (green) and models (pink) respectively. Left panel: comparison of our model to the Local Group dwarfs. Middle panel: comparison of our model to Milky Way dwarfs. Right panel: comparison of our model to M31 dwarfs.}
    \label{fig:t90_hist}
\end{figure}

Despite our overall good agreement, there are interesting distinctions between the modeled and observed $\tau_{90}~-~ \tau_{50}$. Figure \ref{fig:t50_t90} shows that our $\tau_{90}$ values match well with quenching times for Milky Way satellites by \cite{weisz2019}. In Figure~\ref{fig:t90_hist}, we compare the $\tau_{90}$ distributions of \cite{weisz2015} for the Milky Way, M31, and the Local Group as whole. A one sample KS test on the observed distributions with models results in a $p=0.01$ for the Milky Way, $p=0.02$ for the Local Group, and $p=0.12$ for M31 $\alpha=0.02,0.03,0.15$ respectively. Since all $p$--values are less than the corresponding $\alpha$ values none of the modeled distributions come from the same observed distributions of $\tau_{90}$. This may be a sign of either a disagreement between the modeled and observed star formation histories or a sign of inherent scatter in $\tau_{90}$ between halos. However a further exploration of this is beyond the scope of this work.

In contrast, we find a systematic delay of $\tau_{50}$ in our model of about 500~Myrs for the ultra-faint dwarfs. This delay may be due to the lack of molecular hydrogen cooling in our models, delaying the start of star formation until a halo has $v_\mathrm{vir} > v_\mathrm{cooling}$, with $v_\mathrm{cooling}$ chosen to approximate the atomic cooling threshold. While delaying star formation until after the atomic cooling limit does not create the same systematic effect for $\tau_{90}$, it will take the halos longer to form 50\% of their $z~=~0$ stellar populations. We also find the most luminous satellites in our model to have $\tau_{50}\,<\,2 \;Gyrs$. This is a peculiar case since most recent star formation in the Milky Way satellite system took place 3--6~Gyrs based on \cite{weisz2019}.


    
    


\section{Discussion}\label{sec:discussion}
\label{SEC.DISC}

As discussed in the introduction, the well-studied Milky Way satellites are an ideal data set for constraining parameters of Galacticus to best model dwarf galaxies. The initial goal of this study was to build a viable model of the classical dwarfs in the Milky Way and explore predictions for their star formation histories with the standard implementation of Galacticus. However, in addition to successfully modeling the properties of the classical Milky Way satellites, we are also able to match the properties of the more luminous ultra-faint dwarfs. Reproducing the stellar properties of the Milky Way satellites, including the ultra-faint fossil galaxies, was unexpected due to the stochastic star formation processes which govern the evolution of the lowest mass galaxies \citep{Guo+2016}. In addition to the properties at $z\,=\,0$, we also reproduce the star formation histories and quenching times ($\tau_{90}$ vs $\tau_{50}$) of the Local Group dwarfs.\\

Despite the success of Galacticus in modeling the dwarfs, the match between the $z=0$ properties and star formation histories for the classical and brightest ultra-faints breaks down for the dwarfs below $M_\mathrm{V}~\sim-6$). Dwarf galaxy halos modeled with Galacticus cool via atomic processes. As discussed in Section~\ref{subsec:constrained}, we choose the minimum $v_{vir}$ to approximate the atomic cooling cut off during reionization. The lowest mass dwarfs ($M~<~10^8~M_\odot$) initially cooled via the rotational and vibrational transitions of $H_2$ \citep{bromm2019}. The lack of $H_2$ cooling in our model delays the start of star formation in all our dwarfs. Since the majority of stars in the more luminous dwarfs ($M_V~>~-8$) formed when their host halos were above the atomic cooling threshold, we are able to model their properties and star formation histories. In contrast, as the luminosity, and halo mass \citep{Santos+2022}, of the faintest dwarfs decreases, fraction of the stars formed with $v_{vir}~<~v_\mathrm{cooling}$ increases. Since our model does not currently account for gas cooling via $H_2$, we are less able to reproduce the properties and star formation histories of the latter group. In addition, a subset of the faintest dwarfs never reach $v_\mathrm{vir}~>~v_\mathrm{cooling}$. As a result, they remain completely dark in our model, an effect seen by the turnover of the modeled luminosity function at $M_V~>~-4$. The question on whether star formation in halos with masses $<10^8$~M$_\odot$ at reionization (below the atomic cooling limit) is required to reproduce the observed properties of UFDs is still an open question. A robust test of what is the minimum halo mass hosting luminous galaxies has been proposed in \citep{KangR:2019,RicottiPC:2022} and relies on detecting (``ghostly") stellar halos in isolated dwarf galaxies in the Local Group ({\it e.g.,} Leo~A, WLM, IC~1613, NGC~6822). The first results using this new method seem to indicate that halos with masses as low as $10^7$~M$_\odot$ at $z\sim 7$ should be luminous. This is also in agreement with results from DES \citep{Nadleretal:2020} using halo-matching \citep{behroozi2019}. The inclusion of models of $H_2$ into Galacticus, and how it effects our modeling of the faintest dwarfs, will be a subject of parallel work.

Several previous studies of dwarf galaxies have been made using SAMs, several of which have found it challenging to reproduce a broad range of dwarf galaxy properties without significant modification to the SAM. For example, \cite{Lu2017} found that their SAM could not simultaneously produce a good match to the dwarf galaxy mass function and mass-metallicity relation without the introduction of a preventative feedback model which reduced the fraction of baryons accreting into a halo as a function of its mas and redshift of that halo. Similarly, \cite{sommerville2020} found that their SAM predicted gas accretion rates orders of magnitudes higher than those found in the FIRE-2 simulations \citep{2014MNRAS.445..581H,2018MNRAS.480..800H}, and were driven to allow stellar feedback to heat gas surrounding halos and thereby preventing it from accreting at such high rates. While we have not explored gas accretion rates in this paper, we \emph{have} examined the luminosity function and mass-metallicity relation---essentially the diagnostics used by \cite{Lu2017}. Interestingly, we do not find the need for any preventative feedback to simultaneously match both of these quantities.

The reasons for this lack of need for preventative feedback are not immediately clear. While the physics ingredients of the Galacticus SAM are fundamentally very similar to the SAMs of both \cite{Lu2017} and \cite{sommerville2020}, there are differences in the details of the physics models. Additionally, there are differences in the numerical implementations of models (e.g. Galacticus uses an adaptive timestep ODE solver, while the SAMs of \cite{Lu2017} and \cite{sommerville2020} use fixed steps, often with each physical process applied in succession, rather than simultaneously). Identifying the primary cause of the lack of need of preventative feedback in Galacticus is a key question, but one which requires an extensive study that is beyond the scope of this present work. Fortunately, the flexible, modular nature of Galacticus allows the possibility of constructing models which mimic the SAMs of \cite{Lu2017} and \cite{sommerville2020}---this will allow us to explore in detail which physical or numerical choices lead to these different conclusions. We intend to undertake such a detailed study in a follow-up work.

\section{Summary and Conclusions}\label{sec:summary}

In this work, we have modeled the Milky Way satellites using the semi-analytic model Galacticus \citep{galacticus} run on merger trees from a high resolution N-body simulation of a Milky Way analog. Using available astrophysical priors, we tune the gas cooling in halos, star formation and feedback recipes to reproduce the observed luminosity function and the luminosity-metallicity relation of the Milky Way satellites \citep{mcconnachie2012,drlica-wagner2020} and the simulated luminosity functions from the Mint Justice League \citep{Applebaum2021}, and mock observations of FIRE-II \citep{Shipp+2022}.


Our conclusions are as follows.
\begin{itemize}
    \item We reproduce the luminosities and metallicities of the Milky Way satellites down to $M_\mathrm{V}~\sim~-6$. In addition, despite the lack of $H_2$ cooling in our current model, we successfully model the properties of the most luminous of the ultra-faint fossil dwarfs. 
    
    \item When our model is tuned to reproduce the observed luminosity function and luminosity-metallicity relation, we are able to independently reproduce several $z~=~0$ properties of the Milky Way dwarfs, including half light radii, velocity dispersion and mass to light ratios without any additional tuning of the physics.
    
    
    \item In addition to reproducing the observed $z~=~0$ properties of Milky Way dwarfs with $M_\mathrm{V}~<~-6$, our work produces star formation histories that are consistent with observations \citep{weisz2014,weisz2019}. As with the additional $z~=~0$ dwarf properties, this is done without any additional tuning of the baryonic physics. We also find the quenching timescale of our modeled dwarfs are in reasonable agreement with those for the M31 dwarfs.
\end{itemize}

This work shows the ability of Galacticus to reproduce the $z~=~0$ properties and star formation histories of dwarf galaxies down to the luminous ultra-faints, providing a new tool for investigating the astrophysics of star formation and feedback in the lowest mass systems. \\

\noindent The authors acknowledge the University of Maryland supercomputing resources (http://hpcc.umd.edu) made available for conducting the research reported in this paper. We thank Ferah Munshi, Charlotte Christensen, Alyson Brooks of MINT Justice League collaboration and Nondh Panithanpaisal, Nora Shipp, and Robyn Sanderson of FIRE-II collaboration for giving access to their respective luminosity functions.  SW thanks Alessa Ibrahim Wiggins and Natalie Myers for proof reads. MSB and SW thank Minerva for many contributing conversations and encouragement. 

\bibliography{sample631}{}
\bibliographystyle{aasjournal}



\end{document}